\documentclass
[floatfix,superscriptaddress,secnumarabic,amssymb,amsmath,nobibnotes,aps,prd,showkeys,showpacs,nofootinbib,onecolumn,notitlepage,12pt]{revtex4}%
\usepackage{setspace}
\usepackage{color}
\usepackage{amsmath}
\usepackage{amsfonts}
\usepackage{verbatim}
\usepackage{amssymb}
\usepackage{graphicx,bm}
\usepackage{graphicx}
\usepackage{amsmath}
\usepackage{amssymb}
\usepackage{amssymb}
\usepackage{graphicx,bm}
\usepackage{graphicx}
\usepackage[caption=false]{subfig}%
\providecommand{\U}[1]{\protect\rule{.1in}{.1in}}

\newcommand{\be}{\begin{equation}}
\newcommand{\ee}{\end{equation}}

\newcommand{\mincir}{\raise
-3.truept\hbox{\rlap{\hbox{$\sim$}}\raise4.truept\hbox{$<$}\ }}
\newcommand{\magcir}{\raise
-3.truept\hbox{\rlap{\hbox{$\sim$}}\raise4.truept\hbox{$>$}\ }}

\ifx\pdfoutput\relax\let\pdfoutput=\undefined\fi
\newcount\msipdfoutput
\ifx\pdfoutput\undefined\else
\ifcase\pdfoutput\else
\ifx\paperwidth\undefined\else
\ifdim\paperheight=0pt\relax\else\pdfpageheight\paperheight\fi
\ifdim\paperwidth=0pt\relax\else\pdfpagewidth\paperwidth\fi
\fi\fi\fi
\begin{document}
\title{Attractors in $f\left(  Q,B\right)  $-gravity}
\author{Andronikos Paliathanasis}
\email{anpaliat@phys.uoa.gr}
\affiliation{Institute of Systems Science, Durban University of Technology, Durban 4000,
South Africa}
\affiliation{Departamento de Matem\'{a}ticas, Universidad Cat\'{o}lica del Norte, Avda.
Angamos 0610, Casilla 1280 Antofagasta, Chile}

\begin{abstract}
We investigate the asymptotic behavior of the cosmological field equations in
Symmetric Teleparallel General Relativity, where a nonlinear function of the
boundary term is introduced instead of the cosmological constant to describe
the acceleration phase of the universe. Our analysis reveals constraints on
the free parameters necessary for the existence of an attractor that
accurately represents acceleration. However, we also identify asymptotic
solutions depicting Big Rip and Big Crunch singularities. To avoid these
solutions, we must impose constraints on the phase-space, requiring specific
initial conditions.

\end{abstract}
\maketitle

\section{Introduction}

\label{sec1}

Recent cosmological observations \cite{od1,od2,od3} challenge Einstein's
General Relativity (GR). In recent years, many cosmologists have proposed
various modified gravitational models to explain these observations
\cite{Buda,Ferraro,mod1,rt11,rt12,f6,rt1,rt2,rt4,rt5,rt7,rt10,ff3,ff4,ff5,rr5,rr6}%
. Although these models can be examined using numerical techniques, analytical
treatment is necessary to derive constraints on the models' free parameters
and draw conclusions about their cosmological viability.

In gravitational physics, the field equations are nonlinear differential
equations, making the derivation of analytic solutions a challenging task.
Certain approaches employed in the literature for constructing analytic
solutions rely on symmetry analysis \cite{ns1,ns2,ns3,ns4} and the
Painlev\'{e} algorithm \cite{p1,p2,p3}. However, in order to understand the
global behaviour of the physical properties in a given gravitational model, we
can study the behaviour of the field equations in the long run. The analysis
of the asymptotics has been widely used in gravitational theories, yielding
many interesting results \cite{dn1,dn2,dn3,dn4,dn5,dn6,dn7,dn8,dn9,dn10}.

In this work, we investigate the impact of nonlinear boundary corrections in
Symmetric Teleparallel General Relativity (STGR/STEGR) \cite{nester} on
cosmological solutions. STGR is a gravitational theory where the physical
space is described by metric tensor $g_{\mu\nu}$ and the symmetric, flat
connection $\Gamma_{\mu\nu}^{\lambda}$ with the covariant derivative
$\nabla_{\lambda}$ such that $\nabla_{\lambda}g_{\mu\nu}\neq0$ and
$Q_{\lambda\mu\nu}=\nabla_{\lambda}g_{\mu\nu}$ is the nonmetricity tensor. The
nonmetricity scalar $Q$ defined by $Q_{\lambda\mu\nu}$ defines the Lagrangian
function of STGR. Because $Q$ differs from the Ricci scalar $\tilde{R},$
defined by the Levi-Civita connection of the metric tensor, by a boundary term
$B$, STGR is dynamical equivalent with GR. Nevertheless this equivalency is
lost when a nonlinear function of the scalars $Q$ or $B$ are introduced in the
gravitational action. $f\left(  Q\right)  $-theory \cite{lav2,lav3} has been
introduced as alternative dark energy theory where the acceleration of the
universe is attributed to geometrodynamical degrees of freedom.

Recently, the generalized \ $f(Q,B)$-gravity, investigated in a series of
studies \cite{ftc0,ftc1,ftc2}. The boundary term has been introduced before in
the gravitational Action Integral previously within the framework of
teleparallelism \cite{bah} with various applications in cosmological studies,
see for instance \cite{ftb1,ftb2,ftb3,ftb4,ftb5,ftb6,ftb7}. Recently some
black holes solutions in $f\left(  Q,B\right)  $ theory investigated in
\cite{d11}. It was found that in teleparallelism the introduction of a
function with dependency to the boundary term lead to cosmological models
which can explain various eras of the cosmological history \cite{ftb4}. In
this work, we are interest to extend this analysis in the case of symmetric
teleparallel theory of gravity. We focus in the case where the gravitational
Lagrangian is $f\left(  Q,B\right)  =Q+f\left(  B\right)  .~$We introduce the
nonlinear function $f\left(  B\right)  $ to play the role of the dynamical
dark energy. Function $f\left(  B\right)  $ should be nonlinear otherwise STGR
is recovered.

The boundary term introduces higher-order derivatives into the field
equations, which can be attributed to scalar fields \cite{anbound}. These
newly introduced scalar fields play a role in the field equations, imparting
dynamic behavior to dark energy. In the following sections, we utilize
asymptotic analysis to establish constraints on the free parameters of the
gravitational theory and discuss the model's viability based on initial
conditions \cite{dn9,dn10}. The paper is structured as follows.

In Section \ref{sec2} we briefly discuss the basic definitions of STGR and we
introduce the boundary correction term. We consider a
Friedmann--Lema\^{\i}tre--Robertson--Walker (FLRW) geometry and in Section
\ref{sec3} we present the field equations for the gravitational model of our
consideration. Section \ref{sec4} includes the main results of this study
where we present a detailed analysis of the dynamics for the cosmological
field equations. Finally, in Section \ref{sec5} we summarize our results.

\section{Symmetric Teleparallel General Relativity}

\label{sec2}

Consider a four-dimensional (non-Riemannian) manifold characterized by the
metric tensor $g_{\mu\nu}$ and the symmetric and flat connection $\Gamma
_{\mu\nu}^{\kappa}$, defining the covariant derivative $\nabla_{\mu}$.
Furthermore, we assume that the metric tensor and the connection $\Gamma
_{\mu\nu}^{\kappa}$ possess identical symmetries.

Because $\Gamma_{\mu\nu}^{\kappa}$ differs from the Levi-Civita connection it
follows $\nabla_{\lambda}g_{\mu\nu}\neq0$. The tensor field%
\[
Q_{\lambda\mu\nu}\equiv\nabla_{\lambda}g_{\mu\nu}=\frac{\partial g_{\mu\nu}%
}{\partial x^{\lambda}}-\Gamma_{\;\lambda\mu}^{\sigma}g_{\sigma\nu}%
-\Gamma_{\;\lambda\nu}^{\sigma}g_{\mu\sigma}%
\]
is called the nonmetricity tensor and it is the essential for the STGR.

By definition, the connection is symmetric and flat, implying that both the
curvature tensor and the torsion tensor vanish.

From the nonmetricity tensor we can construct the scalar $Q$ as \cite{nester}
\begin{equation}
Q=Q_{\lambda\mu\nu}P^{\lambda\mu\nu}, \label{sd1}%
\end{equation}
which is the Lagrangian function of STGR.

In particular, in STGR the Action Integral is defined as
\begin{equation}
S_{STGR}=\int d^{4}x\sqrt{-g}Q. \label{sd2}%
\end{equation}

Tensor $\ P^{\lambda\mu\nu}$ is the non-metricity conjugate and it is given by
the following expression \cite{nester}%
\begin{equation}
P_{\;\mu\nu}^{\lambda}=-\frac{1}{4}Q_{\;\mu\nu}^{\lambda}+\frac{1}{2}%
Q_{(\mu\phantom{\lambda}\nu)}^{\phantom{(\mu}\lambda\phantom{\nu)}}+\frac
{1}{4}\left(  Q^{\lambda}-\bar{Q}^{\lambda}\right)  g_{\mu\nu}-\frac{1}%
{4}\delta_{\;(\mu}^{\lambda}Q_{\nu)}, \label{defP}%
\end{equation}
and
\begin{equation}
Q_{\mu}=Q_{\mu\nu}^{\phantom{\mu\nu}\nu}~,~\bar{Q}_{\mu}=Q_{\phantom{\nu}\mu
\nu}^{\nu\phantom{\mu}\phantom{\mu}}.
\end{equation}

An equivalent way to write the nonmetricity scalar (\ref{sd1}) is with the use
of the disformation tensor \cite{hh1}%
\begin{equation}
L_{~~\mu\nu}^{\kappa}=\frac{1}{2}\left(  Q_{~\mu\nu}^{\kappa}-Q_{\mu~~\nu
}^{~~\kappa}-Q_{\nu~~\kappa}^{~~\kappa}\right)
\end{equation}
that is,%
\begin{equation}
Q=g^{\mu\nu}\left(  L_{~~\kappa\sigma}^{\kappa}L_{~~\mu\nu}^{\sigma
}-L_{~~\sigma\mu}^{\kappa}L_{~~\nu\kappa}^{\sigma}\right)  \text{.}%
\end{equation}
The disformation tensor depends only on the nonmetricity and defines the
difference of the symmetric and teleparallel connection $\Gamma_{\mu\nu
}^{\kappa}$ with that of the Levi-Civita connection.

Let $\tilde{\Gamma}_{\mu\nu}^{\kappa}$ be the Levi-Civita connection for the
metric tensor $g_{\mu\nu}$, that is $\tilde{\nabla}_{\lambda}g_{\mu\nu}=0$,
and $\tilde{R}$ is the Ricciscalar defined by the Levi-Civita connection. Then
by definition \cite{lav3} $\tilde{R}-Q=B~$where $B$ is a boundary given by the
expression \cite{lav3} $B=-\tilde{\nabla}_{\mu}\left(  Q^{\mu}-\bar{Q}^{\mu
}\right)  $.

Consequently it follows%
\begin{equation}
\int d^{4}x\sqrt{-g}Q\simeq\int d^{4}x\sqrt{-g}\tilde{R}+\text{boundary
terms.}%
\end{equation}
and the theory is equivalent to GR.

The gravitational field equations of STGR are
\begin{equation}
\frac{2}{\sqrt{-g}}\nabla_{\lambda}\left(  \sqrt{-g}P_{\;\mu\nu}^{\lambda
}\right)  +\left(  P_{\mu\rho\sigma}Q_{\nu}^{\;\rho\sigma}-2Q_{\rho\sigma\mu
}P_{\phantom{\rho\sigma}\nu}^{\rho\sigma}\right)  -\frac{1}{2}Qg_{\mu\nu}=0.
\end{equation}
\newline

However, variation of (\ref{sd2}) with respect to the connection leads to the
equations of motion%
\[
\nabla_{\mu}\nabla_{\nu}\left(  \sqrt{-g}P_{~~~~\kappa}^{\mu\nu}\right)  =0.
\]
The equation of motion for the connection is not independent from the field
equations. Indeed, if the field equations are satisfied then, the equation of
motion for the connection is also satisfied.

\subsection{Boundary corrections}

The influence of the boundary term on gravitational phenomena has been
previously investigated, particularly in the context of teleparallel gravity
\cite{bah}.

In this study we consider the modified gravitational Action Integral
\begin{equation}
\hat{S}_{STGR}=\int d^{4}x\sqrt{-g}\left(  Q+f\left(  B\right)  \right)  ,
\label{ac.01}%
\end{equation}
where we introduce a nonlinear function $f$ which depend on the boundary $B$.
Action $\hat{S}_{STGR}$ belongs to the family of $f(Q,B)$ (also known as
$f\left(  Q,C\right)  $)$~$theory \cite{ftc0,ftc1,ftc2}. In our consideration
we assume the dynamical degrees of freedom provided by the correction term
$f\left(  B\right)  $ to play the role of a dynamical dark energy.

The gravitational field equations are%
\begin{align}
0  &  =\frac{2}{\sqrt{-g}}\nabla_{\lambda}\left(  \sqrt{-g}P_{\;\mu\nu
}^{\lambda}\right)  -\frac{1}{2}\left(  Q+B\right)  g_{\mu\nu}+\left(
P_{\mu\rho\sigma}Q_{\nu}^{\;\rho\sigma}-2Q_{\rho\sigma\mu}%
P_{\phantom{\rho\sigma}\nu}^{\rho\sigma}\right) \nonumber\\
&  +\left(  \frac{B}{2}g_{\mu\nu}-\nabla_{\mu}\nabla_{\nu}+g_{\mu\nu}%
g^{\kappa\lambda}\nabla_{\kappa}\nabla_{\lambda}-2P_{~~\mu\nu}^{\lambda}%
\nabla_{\lambda}\right)  f_{,B},
\end{align}
or equivalent%
\begin{equation}
G_{\mu\nu}=T_{\mu\nu}^{f\left(  B\right)  },
\end{equation}
where now $G_{\mu\nu}$ is the Einstein tensor and $T_{\mu\nu}^{f\left(
B\right)  }$ attributes the dynamical degrees of freedom of the boundary term,
that is,
\begin{equation}
T_{\mu\nu}^{f\left(  B\right)  }=\left(  \nabla_{\mu}\nabla_{\nu}-g_{\mu\nu
}g^{\kappa\lambda}\nabla_{\kappa}\nabla_{\lambda}+2P_{~~\mu\nu}^{\lambda
}\nabla_{\lambda}\right)  f_{,B}+\frac{1}{2}\left(  f\left(  B\right)
-Bf_{,B}\right)  g_{\mu\nu}.
\end{equation}

We introduce the scalar field $\zeta=f_{,B}$; thus, the energy momentum tensor
$T_{\mu\nu}^{f\left(  B\right)  }$ becomes%
\begin{equation}
T_{\mu\nu}^{f\left(  B\right)  }=\left(  \nabla_{\mu}\nabla_{\nu}\zeta
-g_{\mu\nu}g^{\kappa\lambda}\nabla_{\kappa}\nabla_{\lambda}\zeta+2P_{~~\mu\nu
}^{\lambda}\nabla_{\lambda}\zeta\right)  +\frac{1}{2}V\left(  \zeta\right)
g_{\mu\nu},
\end{equation}
where $V\left(  \zeta\right)  =\left(  f\left(  B\right)  -Bf_{,B}\right)  .$

Furthermore, the equation of motion for the connection for the Action Integral
(\ref{ac.01}) reads \cite{ftc2}%
\begin{equation}
\nabla_{\mu}\nabla_{\nu}\left(  \sqrt{-g}\zeta P_{~~~~\kappa}^{\mu\nu}\right)
=0\,\,.
\end{equation}
As a result new dynamical variables are introduced by the selection of the
connection. When the latter equation is trivial satisfied, we shall say that
the connection is defined in the coincidence gauge. For more details we refer
the reader to \cite{hh1}.

The gravitational model (\ref{ac.01}) belongs to the family of $F\left(
Q,\tilde{R}\right)  $ theory with Action Integral
\begin{equation}
S_{F\left(  Q,\tilde{R}\right)  }=\int d^{4}x\sqrt{-g}\left(  F\left(
Q,\tilde{R}\right)  \right)  .
\end{equation}
Recall that the $\tilde{R}$ is the Ricci scalar for the Levi-Civita
connection. Hence, in order to derive the gravitational field equations for
the the latter gravitational Action Integral generalized
Gibbons--York--Hawking boundary terms should be introduced. For the $F\left(
\tilde{R}\right)  $-theory the generalized Gibbons--York--Hawking boundary
term is discussed in \cite{boun1}, while for the framework of STEGR the
corresponding generalized Gibbons--York--Hawking boundary term discussed
recently in \cite{boun2}. We remark that for $F\left(  Q,\tilde{R}\right)
=Q+f\left(  \tilde{R}-Q\right)  $, we recover the gravitational model of our
consideration (\ref{ac.01}); see also the disucssion in \cite{ftc1}.

\section{FLRW Cosmology}

\label{sec3}

On very large scales, the universe is isotropic and homogeneous, described by
the spatially flat FLRW geometry with the line element
\begin{equation}
ds^{2}=-N\left(  t\right)  dt^{2}+a^{2}\left(  t\right)  \left(  dx^{2}%
+dy^{2}+dz^{2}\right)  ,
\end{equation}
where $a\left(  t\right)  $ is the scale factor and $H\left(  t\right)
=\frac{1}{N}\frac{\dot{a}}{a}$ is the Hubble function and $N\left(  t\right)
$ is the lapse function.

The FLRW geometry admits six isometries consisted by the three translation
symmetries%
\[
\partial_{x}~,~\partial_{y}~,~\partial_{z}~,
\]
and the three rotations%
\[
y\partial_{x}-x\partial_{y}~,~z\partial_{x}-x\partial_{z}~,~z\partial
_{y}-y\partial_{z}\text{.}%
\]

The requirement for the connection $\Gamma_{\mu\nu}^{\kappa}$ to be symmetric,
flat, and to inherit the symmetries of the background geometry leads to three
distinct families of connections \cite{Heis2,Zhao}. The cosmological field
equations for these three families of connections were derived previously in
\cite{anbound}.

From the three families of connections, $\Gamma_{1}$,~$\Gamma_{2}$ and
$\Gamma_{3}$ we derive the following nonmetricity scalars%
\begin{equation}
Q\left(  \Gamma_{1}\right)  =-6H^{2}%
\end{equation}%
\begin{equation}
Q\left(  \Gamma_{2}\right)  =-6H^{2}+\frac{3}{a^{3}N}\left(  \frac{a^{3}%
\gamma}{N}\right)  ^{\cdot}%
\end{equation}%
\begin{equation}
Q\left(  \Gamma_{3}\right)  =-6H^{2}+\frac{3}{a^{3}N}\left(  aN\bar{\gamma
}\right)  ^{\cdot}.
\end{equation}
where scalars $\gamma$, $\bar{\gamma}$ have been introduced by the
connections. The corresponding boundary functions $B=\tilde{R}-Q$ are%
\begin{equation}
B\left(  \Gamma_{1}\right)  =3\left(  6H^{2}+\frac{2}{N}\dot{H}\right)  ,
\end{equation}%
\begin{equation}
B\left(  \Gamma_{2}\right)  =3\left(  6H^{2}+\frac{2}{N}\dot{H}-\frac{3}%
{a^{3}N}\left(  \frac{a^{3}\gamma}{N}\right)  ^{\cdot}\right)
\end{equation}
and%
\begin{equation}
3\left(  6H^{2}+\frac{2}{N}\dot{H}-\frac{1}{a^{3}N}\left(  aN\bar{\gamma
}\right)  ^{\cdot}\right)  .
\end{equation}

We follow \cite{anbound} and in (\ref{ac.01}) we \ introduce the Lagrangian
multiplier $\tilde{\lambda}$ such that%
\begin{equation}
\hat{S}_{STGR}=\int d^{4}x\sqrt{-g}\left(  Q\left(  \Gamma_{I}\right)
+f\left(  B\right)  +\tilde{\lambda}\left(  B-B\left(  \Gamma_{I}\right)
\right)  \right)  ~,~I=1,2,3.
\end{equation}
The equation of motion for the Lagrange multiplier $\frac{\delta S}{\delta
B}=0$, gives $\lambda=-f_{,B}$. Thus, by replacing the nonmetricity and
boundary scalars in the latter Action Integral and integration by part leads
to the following three point-like Lagrangian functions%
\begin{equation}
L\left(  \Gamma_{1}\right)  =-\frac{6}{N}a\dot{a}^{2}-\frac{6}{N}a^{2}\dot
{a}\dot{\zeta}+Na^{3}V\left(  \zeta\right)  , \label{lan.01}%
\end{equation}%
\begin{equation}
L\left(  \Gamma_{2}\right)  =-\frac{6}{N}a\dot{a}^{2}-\frac{6}{N}a^{2}\dot
{a}\dot{\zeta}-3\frac{a^{3}\dot{\zeta}\dot{\psi}}{N}+Na^{3}V\left(
\zeta\right)  ,
\end{equation}
and%
\begin{equation}
L\left(  \Gamma_{3}\right)  =-\frac{6}{N}a\dot{a}^{2}-\frac{6}{N}a^{2}\dot
{a}\dot{\zeta}+3Na\frac{\dot{\zeta}}{\dot{\Psi}}+Na^{3}V\left(  \zeta\right)
,
\end{equation}
in which $\zeta=f_{,B}$ and $V\left(  \phi,\zeta\right)  =\left(
f-f_{,B}\right)  $, $\dot{\psi}=\gamma$ and $\dot{\Psi}=\frac{1}{\bar{\gamma}%
}$. The field equations follow from the variation of the Lagrangian functions
$L\left(  \Gamma_{I}\right)  $ wrt the dynamical variables. Without loss of
generality in the following we assume $N\left(  t\right)  =1$.

For the first connection, namely $\Gamma_{1}$, the modified Friedmann
equations are%
\begin{align}
-3H^{2} &  =3\dot{\zeta}H+\frac{1}{2}V\left(  \zeta\right)  ,\\
-2\dot{H}-3H^{2} &  =\ddot{\zeta}+\frac{1}{2}V\left(  \zeta\right)  ,
\end{align}
in which the scalar field $\zeta$ satisfy the Klein-Gordon equation%
\begin{equation}
\dot{H}+3H^{2}+\frac{1}{6}V_{,\zeta}=0.
\end{equation}
At this point it is interesting to mention that the latter field equations for
connection $\Gamma_{1}$ are the same with that of the teleparallel $f\left(
T,B_{T}\right)  =T+f\left(  B_{T}\right)  $ model \cite{ftb4}. Hence, for the
connection defined in the coincidence gauge we find an one-to-one
correspondence between the two theories. 

For connection~$\Gamma_{2}$ the cosmological field equations are%
\begin{align}
-3H^{2} &  =3\dot{\zeta}H-\frac{3}{2}\dot{\zeta}\dot{\psi}+\frac{1}{2}V\left(
\zeta\right)  ,\\
-2\dot{H}-3H^{2} &  =\frac{3}{2}\dot{\zeta}\dot{\psi}+\ddot{\zeta}+\frac{1}%
{2}V\left(  \zeta\right)  ,
\end{align}
where the scalar fields satisfy the equations motion%
\begin{align}
\ddot{\zeta}+3H\dot{\zeta} &  =0,\\
6\dot{H}+18H^{2}-9H\dot{\psi}-3\ddot{\psi}+V_{,\zeta} &  =0.
\end{align}

Finally, for the third connection, i.e. $\Gamma_{3}$, the field equations are%
\begin{align}
-3H^{2}  &  =3H\dot{\zeta}+\frac{3}{2}\frac{\dot{\zeta}}{a^{2}\dot{\Psi}%
}+\frac{1}{2}V\left(  \zeta\right)  ,\\
-2\dot{H}-3H^{2}  &  =\frac{V\left(  \zeta\right)  }{2}+\frac{\dot{\zeta}%
}{8a^{2}\dot{\Psi}},
\end{align}
and the equations of motion for the scalar fields are%
\begin{align}
3\dot{\Psi}\ddot{\zeta}+\dot{\zeta}\left(  \dot{\Psi}H-2\ddot{\Psi}\right)
&  =0,\\
6\dot{H}+18H^{2}+V_{,\zeta}-\frac{3}{a^{2}\dot{\Psi}^{2}}\left(  H\dot{\Psi
}-\ddot{\Psi}\right)   &  =0.
\end{align}

Connection $\Gamma_{1}$ is defined in the coincidence gauge, whereas
connections $\Gamma_{2}$ and $\Gamma_{3}$ are defined in the noncoincidence
gauge. ~We emphasize that for connections $\Gamma_{2}$ and $\Gamma_{3}$,
scalars $\psi$ and $\Psi$ play crucial roles in the dynamics evolution. It is
clear that the selection of the connection affects the dynamics of the
gravitational model and there is not a unique selection of connection for the
$f\left(  Q,B\right)  $ in a spatially flat FLRW geometry. However, in the
limit of STGR the above field equations reduce to that of GR. We remark that
in the case where the background geometry, the connection defined in the
noncoincidence gauge is usually considered in modified STGR theories
\cite{Heis2,Zhao}$\,$.

We continue our study with the phase-space analysis of the field equations
corresponding to the three connections. For the potential function $V\left(
\zeta\right)  $ we consider the exponential function $V\left(  \zeta\right)
=V_{0}e^{\lambda\zeta}$ which corresponds to the function
\begin{equation}
f\left(  B\right)  =-\frac{B}{\lambda}\ln\left(  -\frac{B}{\lambda V_{0}%
}\right)  -\frac{B}{\lambda}\text{.}%
\end{equation}
As we shall see in the following, the exponential potential is used to reduce
the dimension of the dynamical system. Similar to the usual analysis performed
in other scalar field theories \cite{dn1}.

\section{Analysis of asymptotics}

\label{sec4}

In the following, we conduct an analysis of the asymptotics for the considered
cosmological model. Specifically, we introduce dimensionless variables and
express the field equations as a set of algebraic-differential equations. We
calculate the stationary points and investigate their stability properties.
Each stationary point corresponds to an asymptotic solution with specific
physical properties. Finally, based on the stability properties, we can
establish constraints on the free parameters of the models and discuss the
initial value problem. This analysis is applied to the three different
families of connections.

In the framework of pure $f\left(  Q\right)  $-cosmology, the phase space
analysis and the evolution of the cosmological parameters investigated in
\cite{af1,af2}. Each family of connection provides a different cosmological
evolution. For the first connection the cosmological models is equivalent to
that of teleparallel $f\left(  T\right)  $-theory \cite{p3}. On the other
hand, the other two families of connections provides always the de Sitter
universe as a future attractor. Scaling solutions are provided by the theory
and they can be related to the matter or radiation epochs \cite{af1}. 

\subsection{Connection $\Gamma_{1}$}

To examine the asymptotic evolution of the field equations for the first
connection, we introduce the new variables
\begin{equation}
~z=\frac{\dot{\zeta}}{H}~,~y=\frac{V\left(  \zeta\right)  }{6H^{2}}%
~,~\lambda=\frac{V_{,\zeta}}{V}~,~\tau=\ln a,
\end{equation}
with inverse transformation%
\begin{equation}
\dot{\zeta}=zH~,~V\left(  \zeta\right)  =6yH^{2}~,~V_{,\zeta}=\lambda
V~,~a=e^{\tau}.
\end{equation}

Thus, the field equations transform into
\begin{align}
\frac{dz}{d\tau}  &  =3\left(  1+z\right)  +y\left(  \lambda\left(
2+z\right)  -3\right)  ,\\
\frac{dy}{d\tau}  &  =y\left(  6+\lambda\left(  2y+z\right)  \right)  ,\\
\frac{d\lambda}{d\tau}  &  =\lambda^{2}z\left(  \Gamma\left(  \lambda\right)
-1\right)  ~,~\Gamma\left(  \lambda\left(  \zeta\right)  \right)
=\frac{V_{,\zeta\zeta}V}{\left(  V_{,\zeta}\right)  ^{2}},
\end{align}
and
\begin{equation}
1+y+z=0. \label{ee.01}%
\end{equation}
Moreover, the equation of state parameter $w_{eff}^{\Gamma_{1}}$ is expressed
as follows%
\begin{equation}
w_{eff}^{\Gamma_{1}}=1+\frac{2}{3}\lambda y.
\end{equation}

The latter dynamical system is equivalent with that studied before in the
framework of teleparallel $f\left(  T,B_{T}\right)  $-theory \cite{ftb4}, thus
the phase-space analysis will be the same. However, for the convenience of the
reader we briefly repeat the analysis. 

For the exponential potential, where $\lambda$ is always a constant and with
the application of the latter constraint equation we reduce the field
equations to the single differential equation%
\begin{equation}
\frac{dz}{d\tau}=\left(  1+z\right)  \left(  6-\lambda\left(  2+z\right)
\right)  .
\end{equation}

The stationary points $A=\left(  z\left(  A\right)  \right)  $ of the latter
equation are two (in the finite and infinity regimes), point $A_{1}=-1$ and
point $A_{2}=\frac{6}{\lambda}-2$. Point $A_{1}$ corresponds to a stiff fluid
solution with $w_{eff}^{\Gamma_{1}}\left(  A_{1}\right)  =1$, while for point
$A_{2}$ we calculate $w_{eff}^{\Gamma_{1}}\left(  A_{2}\right)  =-3+\frac
{2\lambda}{3}$, where it follows that the de Sitter universe is recovered for
$\lambda=3$, and the solution describes acceleration for $\lambda<4$. Finally,
for $\lambda>6$ \ point $A_{1}$ is an attractor while for $\lambda<6$, the
attractor is point $A_{2}$.

\subsection{Connection $\Gamma_{2}$}

We introduce the dimensionless variables%
\begin{equation}
x=\frac{\dot{\psi}}{2H}~,~z=\frac{\dot{\zeta}}{H}~,~y=\frac{V\left(
\zeta\right)  }{6H^{2}}~,~\lambda=\frac{V_{,\zeta}}{V}~,~\tau=\ln a,
\end{equation}
that is,%
\begin{equation}
\dot{\psi}=2xH~,~\dot{\zeta}=zH~,~V\left(  \zeta\right)  =6yH^{2}~,~V_{,\zeta
}=\lambda V~,~a=e^{\tau}.
\end{equation}

In terms of the new variables the field equations are
\begin{align}
\frac{dx}{d\tau}  &  =\frac{1}{2}\left(  3x\left(  y-2z-1\right)
+3x^{2}z+3\left(  1+z\right)  +\left(  2\lambda-3\right)  y\right)  ,\\
\frac{dz}{d\tau}  &  =\frac{3}{2}z\left(  y-1+z\left(  x-1\right)  \right)
,\\
\frac{dy}{d\tau}  &  =y\left(  3\left(  1+y\right)  +\left(  \lambda-3\left(
1-x\right)  \right)  z\right)  ,\\
\frac{d\lambda}{d\tau}  &  =\lambda^{2}z\left(  \Gamma\left(  \lambda\right)
-1\right)  ~,~\Gamma\left(  \lambda\left(  \zeta\right)  \right)
=\frac{V_{,\zeta\zeta}V}{\left(  V_{,\zeta}\right)  ^{2}}.
\end{align}

Friedmann's first equation yields the constraint
\begin{equation}
1+z\left(  1-x\right)  +y=0, \label{ee.02}%
\end{equation}
while the equation of state parameter reads%
\begin{equation}
w_{eff}^{\Gamma_{2}}=y-z\left(  1-x\right)  .
\end{equation}

For the exponential potential, i.e. $\lambda$ is a constant, the stationary
points $B=\left(  B\left(  x\right)  ,B\left(  z\right)  ,B\left(  y\right)
\right)  $ are
\begin{equation}
B_{1}=\left(  x,-\frac{1}{1-x},0\right)  ~,~B_{2}=\left(  1-\frac{\lambda}%
{3},0,-1\right)  .
\end{equation}
\qquad\qquad\qquad

$B_{1}$ describes a family of points with correspond to stiff fluid solutions,
that is, $w_{eff}^{\Gamma_{2}}\left(  B_{1}\right)  =1$. On the other hand,
$B_{2}$ describes the de Sitter universe with $w_{eff}^{\Gamma_{2}}\left(
B_{2}\right)  =-1$.

As far as the stability is concerned the eigenvalues of the two-dimensional
system in the space of variables $\left\{  x,z\right\}  $, around the
stationary points $B_{1}$ are $\left\{  0,6-\frac{\lambda}{1-x}\right\}  $,
while around the point $B_{2}$ are $\left\{  -3,-3\right\}  $. Thus, point
$B_{2}$ is always an attractor, while for the stability properties of $B_{1}$
we should employ the center manifold theorem (CMT).

We introduce the new variable $\bar{z}=z+\frac{1}{1-x}$, such that the
coordinates of points $B_{1}$ to be $B_{1}=\left(  x,0,0\right)  $. Then, we
assume $\bar{z}=h\left(  x\right)  $ in order to determine the center
manifold. In order a stable manifold to exist it should hold $h\left(
x_{1}\right)  =0$ and $\frac{dh}{dx}|_{x\rightarrow x_{1}}=0$. We calculate
$h\left(  x\right)  =\frac{1}{1-x}-h_{0}\left(  1-x-\lambda\right)  $; thus,
points $B_{1}$ describe always unstable solutions.

\subsubsection{Poincare variables}

Because the dynamical variables are not constraint, they can take values at
the infinity. Hence, in order to study the analysis at the infinity we
introduce the Poincare variables%
\[
x=\frac{X}{\sqrt{1-X^{2}-Z^{2}}}~,~z=\frac{Z}{\sqrt{1-X^{2}-Z^{2}}}%
~,~dT=\sqrt{1-X^{2}-Z^{2}}d\tau,
\]
where $\left\{  X^{2},Z^{2}\right\}  \leq1$.

In terms of the new variables the field equations are expressed as
\begin{equation}
\frac{dX}{dT}=F_{1}\left(  X,Z\right)  ~,~\frac{dZ}{dT}=F_{2}\left(
X,Z\right)  , \label{d1}%
\end{equation}
while the equation of state parameter is
\begin{equation}
w_{eff}^{\Gamma_{2}}=-1-2Z\frac{X-\sqrt{1-X^{2}-Z^{2}}}{1-X^{2}-Z^{2}}.
\end{equation}

The stationary points $B^{\infty}=\left(  B^{\infty}\left(  X\right)
,B^{\infty}\left(  Z\right)  \right)  $ at the infinity are%
\begin{equation}
B_{1\pm}^{\infty}=\left(  \pm1,0\right)  ~,~B_{2\pm}^{\infty}=\left(
0,\pm1\right)  ,
\end{equation}
and for $\lambda=3$, there exist the family of points%
\begin{equation}
B_{3\pm}^{\infty}=\left(  X,\pm\sqrt{1-X^{2}}\right)  .
\end{equation}

Stationary points $B_{1\pm}^{\infty}$ describe de Sitter solutions, i.e.
$w_{eff}^{\left(  \Gamma_{2}\right)  }\left(  B_{1\pm}^{\infty}\right)  =-1$,
while the asymptotic solutions at points $B_{2\pm}^{\infty}$ describe Big
Crunch or Big Rip singularities, that is $w_{eff}^{\left(  \Gamma_{2}\right)
}\left(  B_{2\pm}^{\infty}\right)  =\mp\infty$. Similarly, the family of
points $B_{3\pm}^{\infty}$ describe Big Crunch and Big Rip singularities.

As far as the stability is concerned, the eigenvalues for the linearized
system around points $B_{1\pm}^{\infty}$ are $\left\{  0,0\right\}  $, from
where we infer that the stationary points describe unstable solutions. For
points $B_{2\pm}^{\infty}$ the eigenvalues are $\left\{  \pm6,\pm\left(
\lambda-3\right)  \right\}  $, which means that the Big Crunch solution
$B_{2-}^{\infty}$ is an attractor for $\lambda>3$. Finally, for $\lambda=3$
the stability of the points $B_{3\pm}^{\infty}$ depend on the sing of the
dynamical variable $X$.

\begin{figure}[ptb]
\centering\includegraphics[width=1\textwidth]{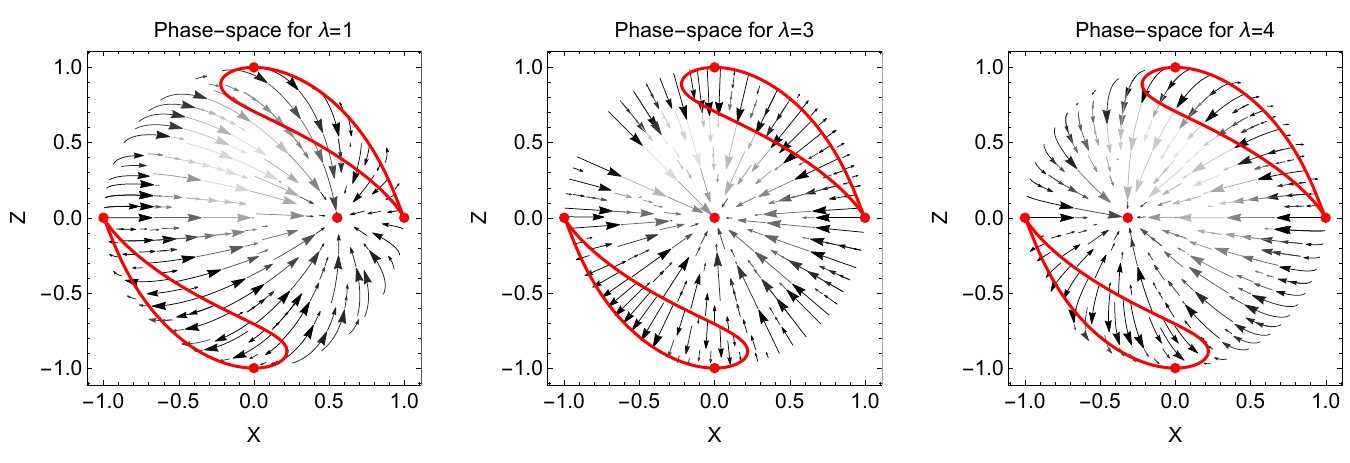}\caption{Phase-space
portraits for the cosmological field equations of connection $\Gamma_{2}$ in
the Poincare variables (\ref{d1}). The phase-space portraits are for
$\lambda=1$, $\lambda=3$ and $\lambda=4$. With dots are the stationary points
and red lines correspond to the family of points $B_{1}$. We observe that for
$\lambda<3$, the unique attractor is the de Sitter solution described by point
$B_{2}$. }%
\label{fig1}%
\end{figure}

\begin{figure}[ptb]
\centering\includegraphics[width=1\textwidth]{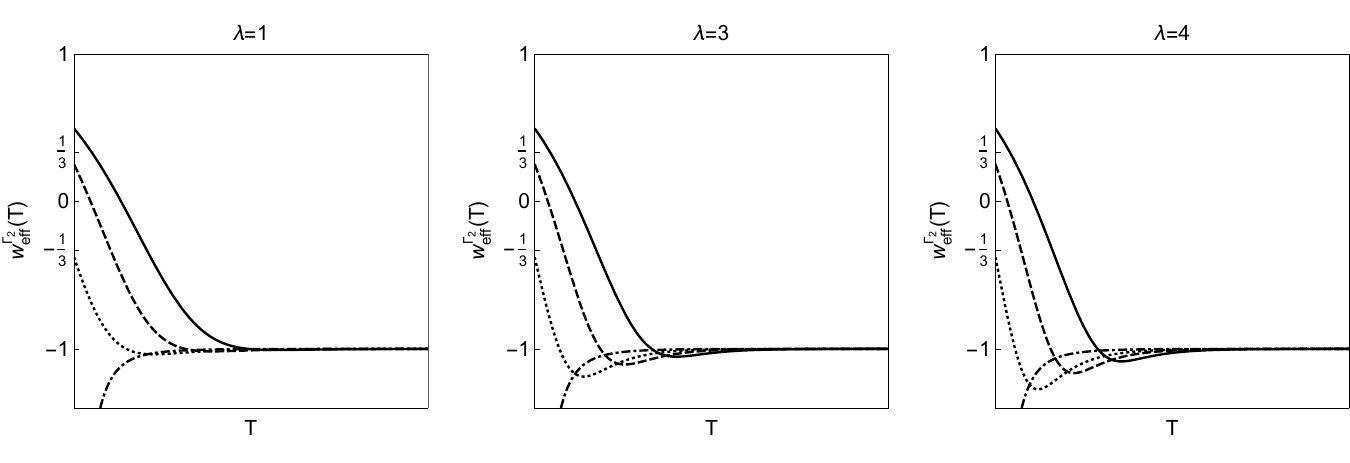}\caption{Qualitative
evolution of the equation of state parameter $w_{eff}^{\left(  \Gamma
_{2}\right)  }$ for different values of $\lambda=\left(  1,3,4\right)  $ and
for various initial conditions $\left(  X_{0},Z_{0}\right)  $. Solid lines are
for $\left(  0.95,0.1\right)  $, dashed lines are for $\left(
0.9,0.19\right)  $, dotted lines are for $\left(  0.8,0.3\right)  $ and
dash-dotted lines are for $\left(  -0.8,0.3\right)  $. }%
\label{fig2}%
\end{figure}

In Fig. \ref{fig1} we present phase-space portraits for the dynamical system,
where it is clear for $\lambda<3$, the unique attractor is the de Sitter
solution described by point $B_{2}$. Moreover, in Fig. \ref{fig2} we present
qualitative evolution of the equation of state parameter for various sets of
initial conditions.

\subsection{Connection $\Gamma_{3}$}

For the set of field equations related to the connection $\Gamma_{3}$ we
introduce the dimensionless variables%
\begin{equation}
\bar{x}=\frac{1}{2a^{2}H\dot{\Psi}}~,~z=\frac{\dot{\zeta}}{H}~,~y=\frac
{V\left(  \zeta\right)  }{6H^{2}}~,~\lambda=\frac{V_{,\zeta}}{V}~,~\tau=\ln a,
\end{equation}
that is,%
\begin{equation}
\frac{1}{\dot{\Psi}}=2a^{2}\bar{x}H~,~\dot{\zeta}=zH~,~V\left(  \zeta\right)
=6yH^{2}~,~V_{,\zeta}=\lambda V~,~a=e^{\tau}.
\end{equation}

The field equations read%
\begin{align}
\frac{d\bar{x}}{d\tau}  &  =\frac{\bar{x}}{2z}\left(  z\left(  1+\bar
{x}\right)  -3+y\left(  3-2\lambda\left(  1-z\right)  \right)  \right)  ,\\
\frac{dz}{d\tau}  &  =3+\left(  3-\bar{x}\right)  z+y\left(  \lambda\left(
2+z\right)  -3\right)  ,\\
\frac{dy}{d\tau}  &  =y\left(  6+\lambda\left(  2y+z\right)  \right)  ,\\
\frac{d\lambda}{d\tau}  &  =\lambda^{2}z\left(  \Gamma\left(  \lambda\right)
-1\right)  ~,~\Gamma\left(  \lambda\left(  \zeta\right)  \right)
=\frac{V_{,\zeta\zeta}V}{\left(  V_{,\zeta}\right)  ^{2}},
\end{align}
and constraint%
\begin{equation}
1+y+\left(  1+\bar{x}\right)  z=0. \label{con1}%
\end{equation}
Moreover, the equation of state parameter is expressed as%
\begin{equation}
w_{eff}^{\Gamma_{3}}=1+\frac{2}{3}\lambda y.
\end{equation}
We remark that for the exponential potential $\lambda$ is always constant.
Hence the stationary points $C=\left(  C\left(  \bar{x}\right)  ,C\left(
z\right)  ,C\left(  y\right)  \right)  $ for the latter algebraic-differential
system are%
\begin{equation}
C_{1}=\left(  0,-1,0\right)  ~,~C_{2}=\left(  0,\frac{2}{\lambda}\left(
3-\lambda\right)  ,1-\frac{6}{\lambda}\right)  .
\end{equation}

Point $C_{1}$ describes a stiff fluid solution, with $w_{eff}^{\Gamma_{3}%
}\left(  C_{1}\right)  =1$. On the other hand, the asymptotic solution at
point $C_{2}$ describes an ideal gas with equation of state parameter
$w_{eff}^{\Gamma_{3}}\left(  C_{2}\right)  =-3+\frac{2}{3}\lambda$. The
stationary point describes acceleration for $\lambda<4$, while the
cosmological constant is recovered for $\lambda=3$.

We make use of the constraint equation (\ref{con1}) and we reduce by one the
dimension of the dynamical system. The eigenvalues of the linearized system
around the stationary points~$C_{1}$ and $C_{2}$ are $\left\{  2,6-\lambda
\right\}  $; $\left\{  \lambda-6,\frac{1}{2}\left(  3\lambda-14\right)
\right\}  $ respectively. Therefore, point $C_{1}$ is a saddle point
when~$\lambda>6$, otherwise is a source; while point $C_{2}$ is an attractor
for $\lambda<\frac{14}{3}$. We remark that when $C_{2}$ describes acceleration
it is always attractor.

\subsubsection{Poincare variables}

For the analysis at the infinity regime we work in the two-dimensional space
defined by the dynamical variables $\left\{  \bar{x},z\right\}  $.

The Poincare variables are defined as%
\[
\bar{x}=\frac{\bar{X}}{\sqrt{1-\bar{X}^{2}-Z^{2}}}~,~z=\frac{Z}{\sqrt
{1-\bar{X}^{2}-Z^{2}}}~,~dT=\sqrt{1-\bar{X}^{2}-Z^{2}}d\tau,
\]
where $\left\{  \bar{X}^{2},Z^{2}\right\}  \leq1$.

The field equations are reduced to the system of the form
\begin{equation}
\frac{d\bar{X}}{dT}=G_{1}\left(  \bar{X},Z\right)  ~,~\frac{dZ}{dT}%
=G_{2}\left(  \bar{X},Z\right)  ,
\end{equation}
and the equation of state parameter is expressed as
\begin{equation}
w_{eff}^{\Gamma_{3}}=1-\frac{2}{3}\lambda\left(  1+\frac{Z\left(  \bar
{X}+\sqrt{1-\bar{X}^{2}-Z^{2}}\right)  }{1-\bar{X}^{2}-Z^{2}}\right)  .
\end{equation}

The stationary points $C^{\infty}=\left(  \bar{X}\left(  C^{\infty}\right)
,Z\left(  C^{\infty}\right)  \right)  $ at the infinity; that is, $1-\left(
\bar{X}\left(  C^{\infty}\right)  \right)  ^{2}-\left(  Z\left(  C^{\infty
}\right)  \right)  ^{2}=0$, are
\[
C_{1\pm}^{\infty}=\left(  0,\pm1\right)  .
\]

We calculate $w_{eff}^{\Gamma_{3}}\left(  C_{1\pm}^{\infty}\right)
=\mp\lambda\infty$. Hence, for $\lambda>0$, $C_{1+}^{\infty}$ corresponds to a
Big Rip singularity, and $C_{1-}^{\infty}$ to a Big Crunch; nevertheless for
$\lambda<0$, $C_{1+}^{\infty}$ corresponds to a Big Crunch singularity, and
$C_{1-}^{\infty}$ to a Big Rip singularity.

The eigenvalues of the linearized system around the stationary points are
$\left\{  \pm2\lambda,0\right\}  $. Because the second eigenvalue is zero, we
employ the CMT and we found that the stationary points does not posses any
submanifold where the solutions are stable. Thus, the stationary points are
saddle points or sources.

\begin{figure}[ptb]
\centering\includegraphics[width=1\textwidth]{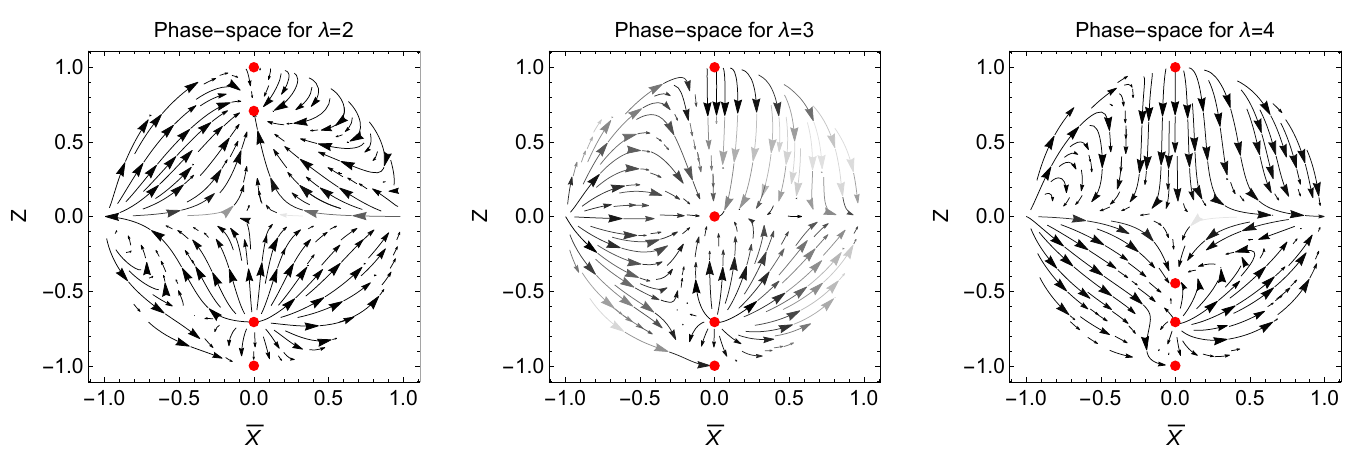}\caption{Phase-space
portraits for the cosmological field equations of connection $\Gamma_{3}$ in
the Poincare variables (\ref{d1}). The phase-space portraits are for
$\lambda=2$, $\lambda=3$ and $\lambda=4$. With dots are the stationary points.
We observe that the unique attractor is point $C_{2}$, while the two
stationary points at the infinity regime are saddle points. }%
\label{fig3}%
\end{figure}

In Fig. \ref{fig3} we present phase-space portraits for the dynamical system
in Poincare variables. Furthermore, in Fig. \ref{fig2} we present qualitative
evolution of the equation of state parameter for various sets of initial conditions.

\begin{figure}[ptb]
\centering\includegraphics[width=1\textwidth]{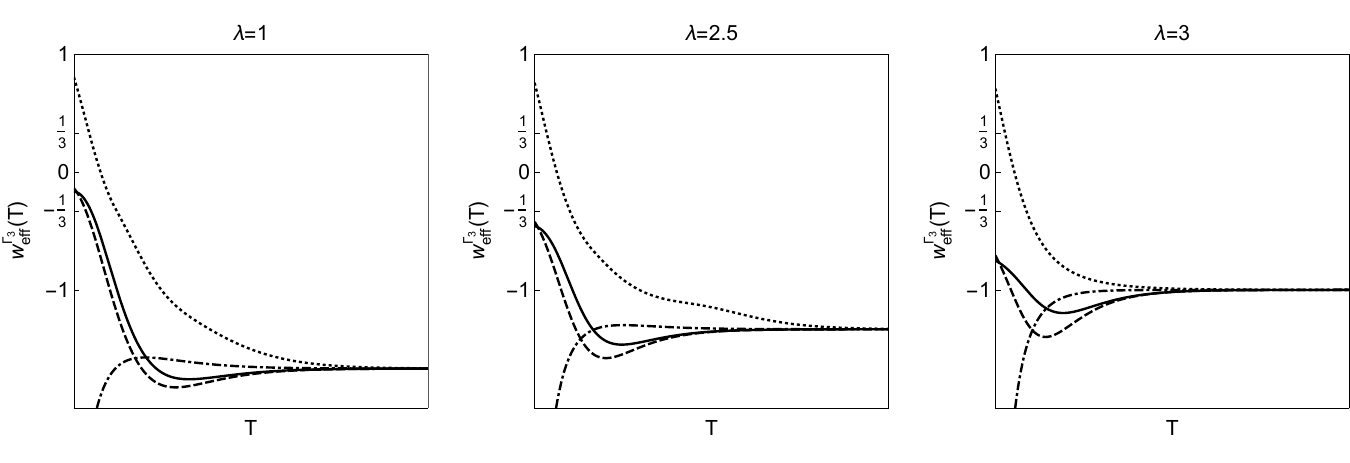}\caption{Qualitative
evolution of the equation of state parameter $w_{eff}^{\left(  \Gamma
_{3}\right)  }$ for different values of $\lambda=\left(  2,2.5,3\right)  $ and
for various initial conditions $\left(  X_{0},Z_{0}\right)  $. Solid lines are
for $\left(  -0.95,0.05\right)  $, dashed lines are for $\left(
-0.8,0.2\right)  $, dotted lines are for $\left(  0.8,-0.2\right)  $ and
dash-dotted lines are for $\left(  0.5,0.7\right)  $. }%
\label{fig4}%
\end{figure}

\section{Conclusions}

\label{sec5}

We conducted a detailed analysis of the asymptotic dynamics for an extension
of STGR in which nonlinear components of the boundary term are introduced in
the gravitational integral. In STGR, the definition of the connection is not
unique, and for the spatially flat FLRW, there are three families of different
connections. Although the selection of the connection does not affect the
gravitational model in STGR when nonlinear terms of the boundary scalar are
introduced, new dynamical degrees of freedom appear. The new degrees of
freedom can be attributed to scalar fields, leading to three different sets of
gravitational field equations, corresponding to the families of connections.

For the three different models, we employed dimensionless variables and
determined the stationary points. We reconstructed the asymptotic solutions at
the stationary points and investigated their stability properties. The results
are summarized in Table \ref{tab1}. For connection $\Gamma_{1}$ the field
equations admit two stationary points, which describe scaling solutions. The
one point correspond to the stiff fluid solution and scale factor $a\left(
t\right)  =a_{0}t^{\frac{1}{3}}$; on the other hand, the second point
describes a scaling solution with scale factor $a\left(  t\right)
=a_{0}t^{\frac{1}{\lambda-3}}$. The second solution describes acceleration for
$\lambda<4$, and the de Sitter universe is recovered for $\lambda=3$.

For the second connection, namely $\Gamma_{2}$, the de Sitter universe exist
as a future attractor for arbitrary value of parameter $\lambda$. Moreover,
the stiff fluid solution exist, while two de Sitter points which can describe
the early acceleration phase of the universe appear. Moreover, for this
connection there exist stationary point which describe Big Rip or Big Crunch
singularities. The Big Crunch singularity is stable when the parameter
$\lambda>3$. Thus, in this case, for the unique attractor to be the de Sitter
solution, it follows that $\lambda\leq3$.

Moreover, for the third connection, $\Gamma_{3}$ we recover the two stationary
points of connection $\Gamma_{1}$, however, unstable Big Rip singularities
appear. For the third connection the unique attractor describes an accelerated
universe for $\lambda<\frac{14}{3}$. time acceleration phase of the universe.

In the previous Sections we have considered the vacuum case. Let us now
introduce a pressureless fluid minimally coupled to gravity, to describe the
dark matter component of the universe. In the presence of the matter source,
the modified first Friedmann's equation has nonzero rhs. Consequently, the rhs
of equations (\ref{ee.01}), (\ref{ee.02}) and (\ref{con1}) are nonzero.
Specifically, they are equation with the energy density $\Omega_{m}$ for the
matter source. Because of the new variable the phase-space has an extra
dimension which means that new stationary points may exist. Indeed, for the
first connection it appears the new point $A_{m}=\left(  z\left(
A_{m}\right)  ,y\left(  A_{m}\right)  \right)  $ with coordinates
$A_{m}=\left(  -\frac{3}{\lambda},-\frac{3}{2\lambda}\right)  $, where
$\Omega_{m}\left(  A_{m}\right)  =1-\frac{9}{2\lambda}$ and $w_{eff}\left(
A_{m}\right)  =0$. This point corresponds to a tracking solution where the
geometric dark energy fluid tracks the dark matter. For connection $\Gamma
_{2}$, there appears the new stationary point $B_{m}=\left(  x\left(
B_{m}\right)  ,z\left(  B_{m}\right)  ,y\left(  B_{m}\right)  \right)  $ with
coordinates $B_{m}=\left(  1,0,0\right)  $, and $\Omega_{m}\left(
B_{m}\right)  =1$, $w_{eff}\left(  B_{m}\right)  =0$. The asymptotic solution
at $B_{m}$ describes a universe dominated by the dark matter. Finally, for the
third connection we find the two extra stationary points $C_{m}=\left(
\bar{x}\left(  C_{m}\right)  ,z\left(  C_{m}\right)  ,y\left(  C_{m}\right)
\right)  $ and coordinates $C_{m}^{1}=\left(  0,-\frac{3}{\lambda},-\frac
{3}{2\lambda}\right)  $, $C_{m}^{1}=\left(  \frac{14-3\lambda}{30},-\frac
{10}{3\lambda},-\frac{4}{3\lambda}\right)  $. Point $C_{m}^{1}$ has the same
physical properties with that of $A_{m}$ for connection $\Gamma_{1}$. On the
other hand, the asymptotic solution at point $C_{m}^{2}$ describes a scaling
solution with $w_{eff}\left(  C_{m}^{2}\right)  =\frac{1}{9}$, and $\Omega
_{m}\left(  C_{m}^{2}\right)  =\frac{4}{3}-\frac{56}{9\lambda}$.\ It is
important to mention that the stability properties of the stationary points
may change in the presence of the matter source. However, such analysis
extends the scopus of this work and will be studied elsewhere.

The above results holds for the exponential scalar field potential , where
parameter $\lambda=\frac{V_{,\zeta}}{V}$ is always a constant function. For a
general functional form of potential $V\left(  \zeta\right)  $, parameter
$\lambda$ is dynamical and three different dynamical systems which we studied
before, are modified by include in all cases the equation of motion for
parameter $\lambda$, that is, the differential equation
\begin{equation}
\frac{d\lambda}{d\tau}=\lambda^{2}z\left(  \Gamma\left(  \lambda\right)
-1\right)  \text{.}%
\end{equation}
For $\lambda=\lambda_{0}$, such that $\lambda_{0}\left(  \Gamma\left(
\lambda_{0}\right)  -1\right)  =0$, we recover the asymptotic solutions for
the exponential potential. In this limit the scalar field potential dominated
by the exponential function. However, there is a new family of solutions, i.e.
stationary points where $z=0$. Then it is easy to see that for arbitrary value
of $\lambda$, for connection $\Gamma_{1}$ we recover point $A_{2}$, for
$\lambda=3$. For the second connection we derive point $B_{2}$, while for the
third connection we get point $C_{2}$ for $\lambda=3$. Hence, the
consideration of the exponential function provides all the possible families
of asymptotic solutions. However, it is important to mention that the
stability properties change, since function $\Gamma\left(  \lambda\right)  $
is introduced in the eigenvalues. 

Nevertheless, it is not obvious from this analysis if this theory can describe
other eras of the cosmological history and if it solves the $H_{0}$-tension.
In a future study, we plan to investigate this specific problem in the context
of $f\left(  Q,B\right)  $-gravity.%

\begin{table}[tbp] \centering
\caption{Asymptotic solutions in STGR with boundary correctons}%
\begin{tabular}
[c]{cccc}\hline\hline
\textbf{Point} & $\mathbf{w}_{eff}$ & \textbf{Acceleration} &
\textbf{Stability}\\\hline
&  &  & \\
\multicolumn{4}{c}{\textbf{Connection} $\mathbf{\Gamma}_{1}$}\\\hline
$A_{1}$ & $1$ & No & Stable$~\lambda>6$\\
$A_{2}$ & $-3+\frac{2}{3}\lambda$ & $\lambda<4$ & Stable$~\lambda<6$\\
&  &  & \\
\multicolumn{4}{c}{\textbf{Connection} $\mathbf{\Gamma}_{2}$}\\\hline
$B_{1}$ & $1$ & No & Unstable\\
$B_{2}$ & $-1$ & Yes & Stable\\
$B_{1\pm}^{\infty}$ & $-1$ & No & Unstable\\
$B_{2\pm}^{\infty}$ & $\mp\infty$ & Big Rip $B_{2+}^{\infty}$ & $B_{2-}%
^{\infty}~$Stable $\lambda>3$\\
$B_{3\pm}^{\infty}~\left(  \lambda=3\right)  $ & $\mp\infty$ & Yes & Stable\\
&  &  & \\
\multicolumn{4}{c}{\textbf{Connection} $\mathbf{\Gamma}_{3}$}\\\hline
$C_{1}$ & $1$ & No & Unstable\\
$C_{2}$ & $-3+\frac{2}{3}\lambda$ & $\lambda<4$ & Stable $\lambda<\frac{14}%
{3}$\\
$C_{1\pm}^{\infty}$ & $\mp\lambda\infty$ & Big Rip & Unstable\\\hline\hline
\end{tabular}
\label{tab1}%
\end{table}%

\begin{acknowledgments}
This work was supported by the UNC VRIDT through Resoluci\'{o}n VRIDT No.
096/2022 and Resoluci\'{o}n VRIDT No. 098/2022. AP thanks the support of
National Research Foundation of South Africa.
\end{acknowledgments}

\end{document}